# Feature-Driven Survey of Physical Protection Systems

Bedir Tekinerdogan[1], Kaan Özcan[2], Sevil Yağız[2], İskender Yakın[2]

[1]Wageningen University & Research, Information Technology, The Netherlands
`bedir.tekinerdogan@wur.nl`
[2]Aselsan A.Ş., Turkey
`syagiz@aselsan.com.tr,mkozcan@aselsan.com.tr,iyakin@aselsan.com.tr`

**Abstract.** Many systems nowadays require protection against security or safety threats. A physical protection system (PPS) integrates people, procedures, and equipment to protect assets or facilities. PPSs have targeted various systems, including airports, rail transport, highways, hospitals, bridges, the electricity grid, dams, power plants, seaports, oil refineries, and water systems. Hence, PPSs are characterized by a broad set of features, from which part is common, while other features are variant and depend on the particular system to be developed. The notion of PPS has been broadly addressed in the literature, and even domain-specific PPS development methods have been proposed. However, the common and variant features are fragmented across many studies. This situation seriously impedes the identification of the required features and likewise the guidance of the systems engineering process of PPSs. To enhance the understanding and support the guidance of the development of PPS, in this paper, we provide a feature-driven survey of PPSs. The approach applies a systematic domain analysis process based on the state-of-the-art of PPSs. It presents a family feature model that defines the common and variant features and herewith the configuration space of PPSs.

**Keywords:** Physical Protection Systems, Systems Engineering, Feature Modeling

## 1 Introduction

Many systems nowadays require protection against security or safety threats. A physical protection system (PPS) integrates people, procedures, and equipment for the protection of assets or facilities [4][16][12]. PPSs have targeted the protection of various systems including airports, rail transport, highways, hospitals, bridges, the electricity grid, dams, power plants, seaports, oil refineries, and water systems. Designing effective PPSs requires careful consideration of the requirements and the resources to provide the protection that is needed. Without a proper assessment and design, valuable resources on unnecessary protection might be wasted or, worse yet, fail to provide adequate protection at critical points of the facility. To avoid both limitations and risks, several PPS have been proposed in the literature to design and analyze PPSs to realize the envisioned objectives. In general, a PPS provides *deterrence*, *detection*, *delay,* and *response* measures to protect against an adversary's attempt to complete a malicious act. Each of these elements has, on its turn, many different features that characterize particular PPSs.

The notion of PPS has been broadly addressed in the literature and even domain-specific PPS development methods have been proposed. However, the common and variant features are fragmented across many studies. This situation seriously impedes the identification of the required features and likewise the guidance of the systems engineering process of PPSs.

To enhance the understanding and support the guidance of the development of PPS, in this paper, we provide a feature-driven survey of PPSs. The approach applies a systematic domain analysis process based on the state-of-the-art of PPSs and presents a family feature model that defines the configuration space of PPSs.

The remainder of the paper is organized as follows. Section 2 presents the background on PPS. Section 3 presents the overall approach and the domain analysis process. Section 4 presents the derived feature model and the survey of PPSs. Section 5 illustrates the guidelines for configuring a PPS based on the family feature diagram. Section 6 presents the related work, finally, section 7 concludes the paper.

## 2    Physical Protection Systems Process

Developing PPSs requires the understanding and application of multiple disciplines and as such requires a systems engineering approach [5]. The traditional systems engineering lifecycle process is often presented as a V-model [7][8], which is a general purpose-process that can be applied for developing various systems. Since it does not directly take into account the domain-specific concerns, several domain-specific methods have been proposed for developing PPSs. Figure 1 shows the top-level activities for a typical PPS design process that we have modeled in our earlier study [12]. In essence, the PPS process consists of three key activities that include determining the PPS objectives, designing the PPS, and evaluating the PPS. Each of these activities can be further refined; the top-level activities of these processes are also shown in  Figure 1.

Determining the PPS objectives includes the facility characterization, the threat definition, and the target definition that needs to be protected. Designing PPS focuses on three activities, detection, delay and response. The resulting PPS design should meet the facility's defined objectives and operational, safety, legal, and economic constraints. The final step in the PPS lifecycle is the evaluation of the design PPS. Several techniques can be distinguished here, including Path Analysis, Scenario Analysis, and System Effectiveness Analysis [3]. For a detailed analysis of these process activities, we refer to [2][3][4][9].

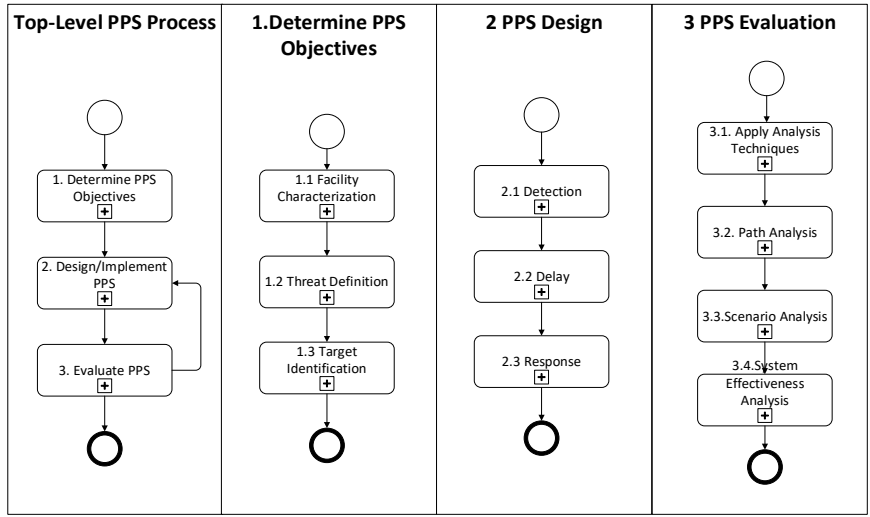

**Figure 1.** PPS Design Process activities (adopted from: [12])

## 3  Domain Analysis

*Domain analysis* can be defined as identifying, capturing, and organizing domain knowledge about the problem domain to make it reusable when creating new systems [11]. A domain is defined as an area of knowledge or activity characterized by a set of concepts and terminology understood by practitioners in that area. Figure 2 represents the domain analysis process that we have followed to model PPSs.

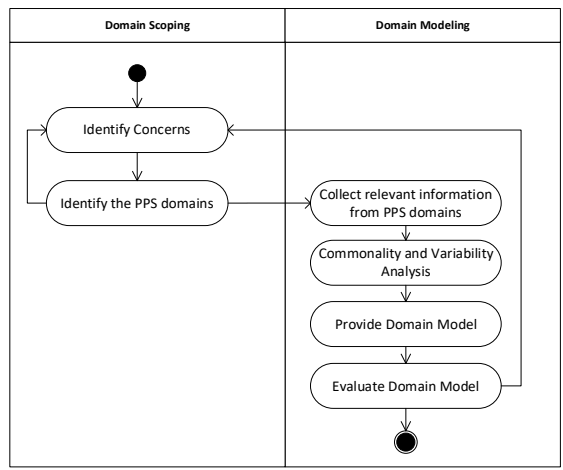

**Figure 2.** Domain Analysis Process

Conventional domain analysis methods generally consist of the activities *Domain Scoping* and *Domain Modeling*: *Domain Scoping* identifies the domains of interest, the stakeholders, and their goals, and defines the scope of the domain. *Domain Modeling* is the activity for representing the domain, or the *domain model*. Typically a domain model is formed through a commonality and variability analysis to concepts in the domain. A *domain model* is used as a basis for engineering components intended for use in multiple applications within the domain. In the context of domain analysis, variability modeling plays an important role and is one of the key activities in the product line engineering. Since its introduction, feature modeling has been widely adopted by the software product line community and a number of extensions have been proposed. A feature is a system property that is relevant to some stakeholders and is used to capture commonalities or discriminate between systems. A feature model is a model that defines features and their dependencies. Feature models are usually represented in feature diagram (or tables). A feature diagram is a tree with the root representing a concept (e.g., a software system), and its descendent nodes are features. Relationships between a parent feature and its child features (or sub-features) include *mandatory*, *optional*, *or,* and *alternative* features. A feature diagram depicts the common and variant features of a particular system and thus the configuration space. A feature constraint further restricts the possible selections of features to define configurations. The most common feature constraints are *requires* and *excludes* constraints.

**Table 1.** Adopted list of studies on Physical Protection Systems

| |
|---|
| ML. Garcia. Vulnerability Assessment of Physical Protection Systems. Amsterdam: Elsevier Butterworth-Heinemann; 2006. [3] |
| ML. Garcia. The Design and Evaluation of Physical Protection Systems. 2nd ed. Amsterdam: Elsevier Butterworth-Heinemann; 2008. [4] |
| L. Fennelly. Effective Physical Security, Fifth Edition (5th. ed.). Butterworth-Heinemann, USA, 2016. [9] |
| J.D. Williams, Physical Protection System Design and Evaluation, IAEA-CN-68/29, Vienna, 10–12 November 1997. [16] |
| IAEA, Handbook on the Physical Protection of Nuclear Material and Facilities, IAEA-TECDOC-127, March 2000. [6] |

For the PPS methods we have primarily consulted the methods of the sources as shown in Table 1. The goal of domain modelling is usually to support the understanding of the common and variant features, but also to provide guidance for the configuration of specific PPSs. In this paper, we adopt both goals, that is, with the provided domain model we aim to support both the understandability of the PPS configuration space, and guide the development of PPSs. In this context, the resulting domain model will be used as part of a product line engineering activity that focuses on the large scale reuse based development of PPSs. This idea is shown in Figure 3. The domain engineering in this process will result in a reusable asset base that is then used in the application engineering to develop various PPSs. The domain model that we describe in this paper is part of the reusable asset-based.

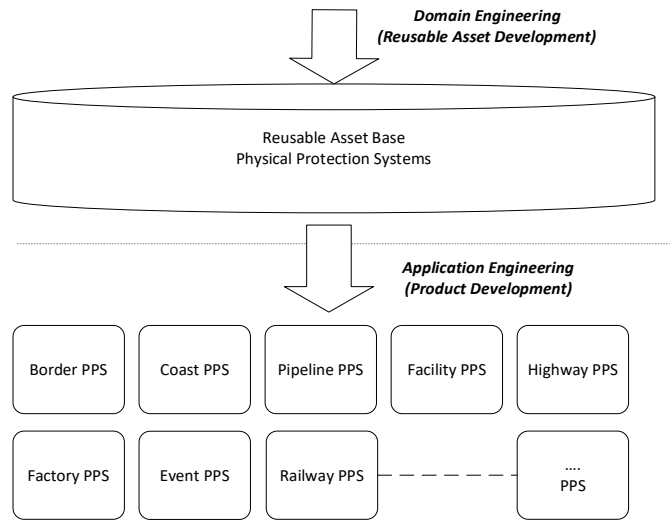

**Figure 3.** Overall Scope for PPS

## 4 Feature Model PPS

### 4.1 Top-Level Feature Diagram

Figure 4 shows the top-level feature diagram for PPS. The PPS has three key mandatory features including, target, threat, and PPS Element.

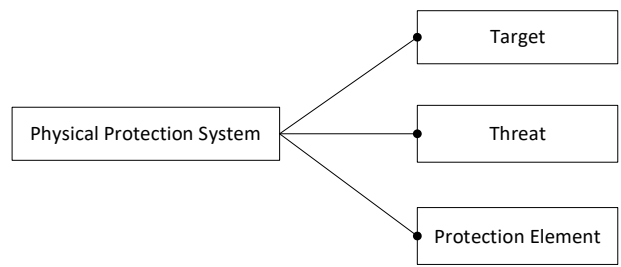

**Figure 4.** Top-Level Feature Diagram for PPS

Each of these features has sub-features, which we will explain in the following subsections. *Target* represents the potential target of an adversary action. *Threat* defines the adversarial properties on a facility. The *protection elements* define the elements for deterrence, detection, delay, and response.

## 4.2 Target

Figure 5 shows the feature diagram for the target feature which has two sub-features target domain and target properties. Since we have adopted a broad scope, the feature target domain relate to various different domains. The list is also open-ended and more domains can be added. The selection of the domain can impact the selection of the other features of the PPS feature diagram.

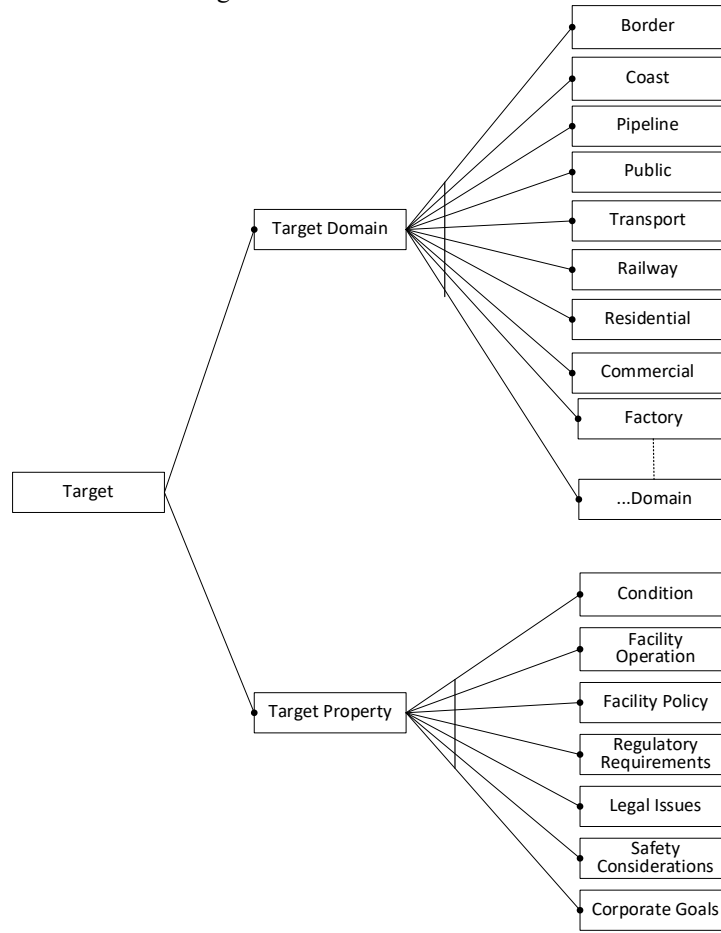

**Figure 5.** Feature Diagram for PPS Domains

For designing the PPS, it is important to know the key characteristics of the target. This requires an understanding of the surrounding environment of the target, the constraints, expected performance, operations, and a clear idea what needs to be protected. Neglecting a proper characterization will lead to either overprotect a target or fail to adequately protect the target. Overprotection will lead to unnecessary waste of resources and increase the cost. Inadequate protection will leave the target vulnerable to potential adversarial attacks. Many different characteristics can be identified which can be grouped as sub-features of the feature *target property*. Conditions represent the infrastructure details such as target boundary, location, access points, existing physical protection features, and other infrastructure details. Operations defines the adopted processes for the target, such as operating conditions (working hours, off hours, emergency operations), and employees' types and numbers. Policies and procedures define the written and unwritten policies and procedures used. Regulatory requirements define the regulatory authorities, such as, local fire department, safety and health regulators, and federal agencies. Legal issues define the legally defined issues such as cover liability, privacy, access for the disabled, labor relations, employment practices, proper training for guards, the failure to protect, and excessive use of force by guards. Safety considerations are issues related to safety. Corporate goals and objectives define the goals and objectives of the corporation or facility regarding protection.

### 4.3 Threat

Figure 6 shows the feature diagram for threats. Before discussing the specific design elements for realizing the protection of the physical system a proper threat definition must be considered. In this context the notion of design basis threat (DBT) is often used to establish the expected threat to a facility and likewise to support the decision-making for the selection of the PPS design elements. The feature diagram of Figure 6 shows the related features for threats and thereby can be useful to define the DBT. A threat is defined as an entity with motivation, intention and capability to commit a malicious act. Different threat sources can be distinguished including hackers, criminals, terrorists, extremists and natural threat sources. Adversaries can be grouped into different groups including insiders, outsiders, and outsiders collaborating with insiders.

Different classes of adversaries include outsiders, insiders, and outsiders working in collusion with insiders. The range of tactics of adversaries includes force, stealth, deceit or any combination of these. As defined by Garcia [4], deceit is the attempted defeat of a security system by using false authorization and identification; force is the overt, forcible attempt to overcome a security system; and stealth is any attempt to defeat the detection system and enter the facility covertly. Different adversary actions can be distinguished including theft, industrial espionage, sabotage, extortion, blackmail and kidnapping. The motivation of an adversary can be ideological, economic, personal or irrational. Finally adversaries may have different capabilities including the number of adversaries, the used weapons, the equipment and tools, the transportation means, the technical experience and insider assistance.

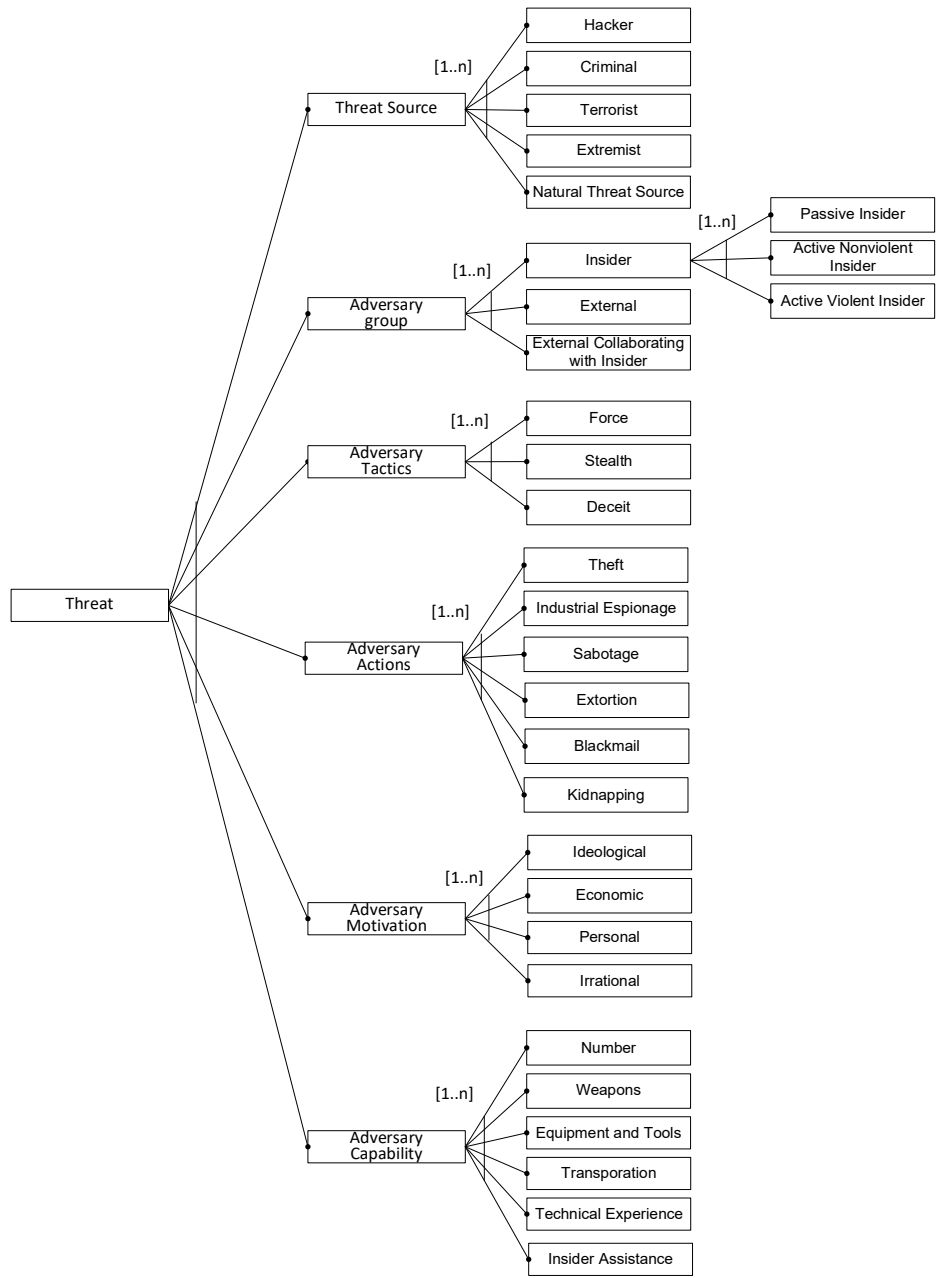

**Figure 6.** Feature Diagram for PPS Threats and Risks

## 4.4 Protection Element

After the characterization of the facility and the analysis and definition of the threats in the DBT the PPS can be designed. The PPS design should meet the defined objectives and operational, safety, legal, and economic constraints of the facility. Figure 7 shows the feature diagram for the PPS design elements. We can identify four basic steps in the PPS design, that is, deterrence element, detection element, delay element and response element. Since deterrence and delay elements are largely referring to similar components we have put them under the same feature root.

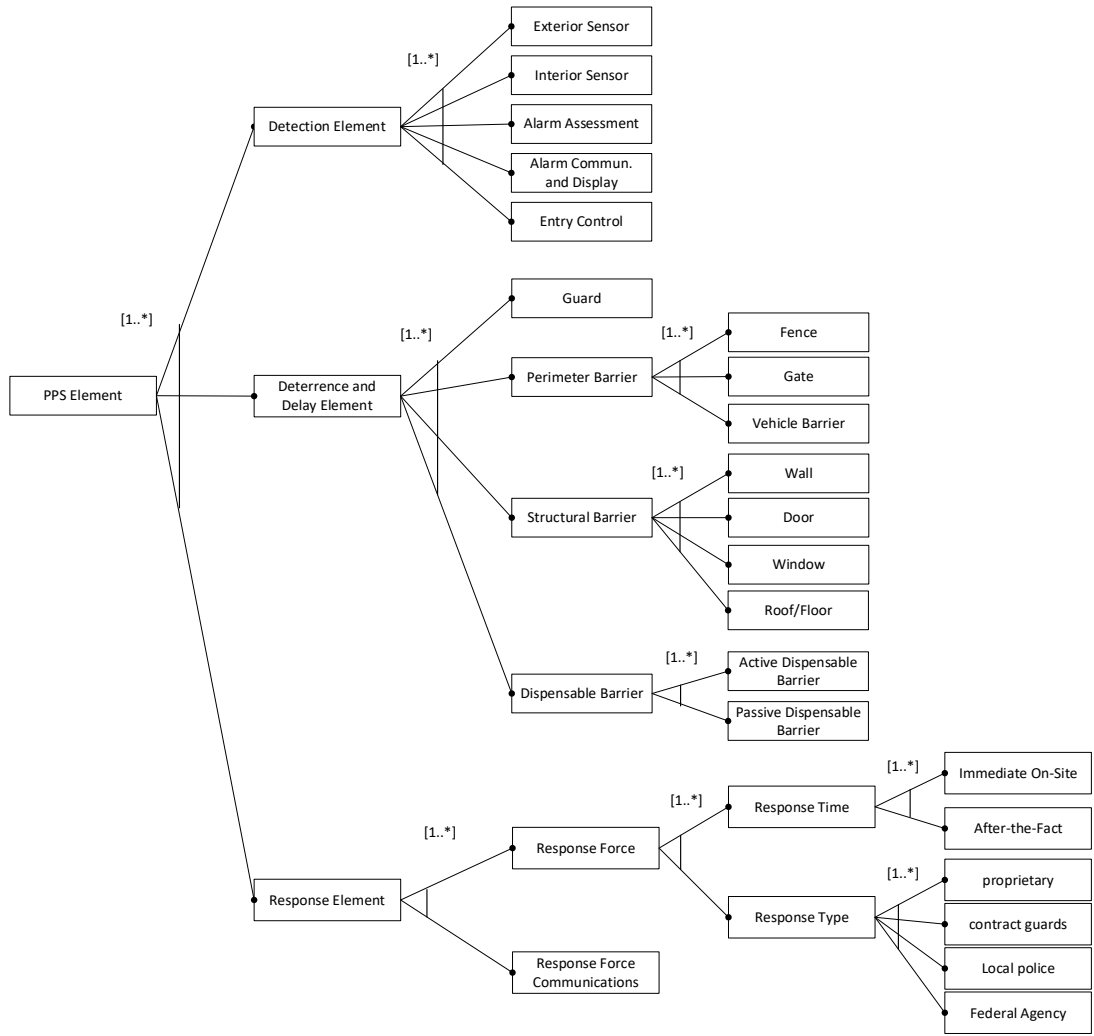

**Figure 7.** Feature Diagram for PPS elements

Detection includes the discovery of an adversary action that is covert or overt. Various detection elements can be identified including exterior sensor, interior sensor, alarm assessment, alarm communication and display, and entry control. Exterior sensors are used in an outdoor environment, while interior sensors are those used inside facilities. Alarm assessment defines the assessment of an alarm that is initiated after detection of an adversary action by interior or exterior sensors. Alarm communication defines the communication of the alarm assessment and the corresponding decision. The assessed alarm can be defined as a false alarm or a real adversary action. Detection elements also include entry control, which allows entry to authorized personnel only, and detects unauthorized personnel or material's attempted entry. In general, the detection function's effectiveness is defined by the probability of sensing adversary action and the time required for reporting and assessing the alarm.

Deterrence and delay elements include the sub-features of guards, perimeter barriers, structural barriers, and dispensable barriers. Perimeter barriers refer to natural barriers or built fortifications, and form the outermost protective layer of a PPS. Some of the key features include fences, gates, and vehicle barriers. Dispensable barriers include the elements that are deployed only when necessary, that is, during an adversary attack. Dispensable barriers can be either active or passive. Active dispensable barriers are controlled by a corresponding command control mechanism and thus can, on command, stop or delay an adversary from accomplishing the objective. In contrast to active dispensable barriers, passive dispensable barriers do not require any command and control system, and likewise reduce the overall cost.

The response function is triggered after the detection of an adversary attack, and includes responding personnel and the communications system that is used. The response force can be either on-site or after-the-fact. The type of response force can be proprietary or contract guards, local and state police, and, or federal agencies.

## 5    Application Feature Model

The PPS family feature model that we have defined in the previous section can be used for two purposes. On the one hand it can be used to support the understanding of the PPS domain, on the other hand it can be used to guide the design of the PPS system. Figure 8 shows the conceptual model representing the PPS family feature model's relation and the PPS application feature model.

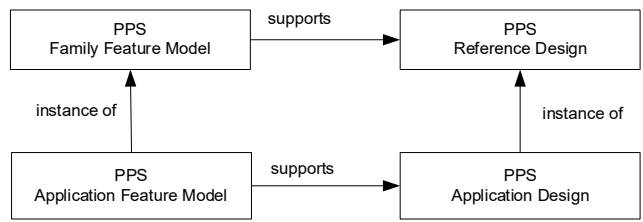

**Figure 8.** Application Feature Diagram

Further, the figure also shows the relation to the PPS reference design and PPS application design. The PPS reference design can also be considered as a product line engineering architecture [13][14][16]. As we have discussed before, the design elements related to deterrence, detection, delay and response elements. The feature model that we have discussed in the previous section represents a family feature model since it is defined for a family of PPSs, that is, it defines the configuration space of the potential PPSs. A specific configuration of the family feature diagram represents a PPS application feature model.

## 6     Related Work

Several studies have provided a detailed overview of PPS methods describing the key elements [2][3][4][6]. Unfortunately, none of these have provided a comprehensive overview of a commonality and variability perspective. We have provided a feature modelling approach that explicitly shows the common and variant features for developing PPSs.

In this paper, we primarily focused on the features of PPSs and did not consider the process activities. In our earlier studies, we did focus on modelling the process of PPS and integrating this with a product line engineering process [12][13][15][16]. This is useful in case multiple PPS systems need to be developed based on systematic reuse. Product line engineering provides a large scale systematic reuse approach, but the presented methods are general purpose and do not explicitly consider PPS concerns. The integrated process as defined in [12] can be used for product line engineering of PPSs and thus reduce the cost of development, reduce the time-to-market, and increase the quality. The family feature model presented in this paper is complementary and can be integrated into the PLE process's domain engineering process.

 Related to PPS and perhaps part of a PPS is an intrusion detection system (IDS), which has also been extensively discussed in the literature. An IDS monitors a network or system for malicious activity or policy violations [1][10]. In case of an intrusion or violation activity, this is typically reported either to an administrator or collected centrally using a security information and event management system. The latter combines outputs from multiple sources and uses alarm filtering techniques to distinguish malicious activity from false alarms. In the PPS, the detection functionality largely uses the techniques as proposed by IDS and IDPS (intrusion detection and prevention system). A related study in this context could be the detailed feature modelling for IDS.

## 7     Conclusion

In this paper, we have provided a domain model for PPS that can be used to characterize many different PPSs. The feature model has been derived after a thorough domain analysis on PPSs and, as such, builds on existing literature. The distinguishing factor of this study is the explicit focus on the commonality and variability analysis. Feature modelling appeared to be a very useful approach for this purpose. Additional studies can be carried out to further detail the sub-features. Using a case study we could easily characterize a real industrial PPS case study to validate the family feature model. We will carry out further case studies to enhance the feature model further in the future. Besides, the ontological support and the focus on understanding the key features, the family feature model can also be used in the threats definition process which results in a design basis threat document. Further, the PPS family feature model can be embedded in a product line engineering processes to support the design of a PPS architecture. We consider this also part of the future work.